\documentclass[prd,showpacs, twocolumn,amsmath,amssymb]{revtex4}

\usepackage{amssymb}
\usepackage{graphicx}
\usepackage{dcolumn}
\usepackage{bm}
\usepackage{amsfonts}
\usepackage{keyval,graphicx}
\usepackage{textcomp,wasysym}

\newcommand{\etap}{\eta^{\prime}}

\newcommand{\pip}{\pi^{+}}
\newcommand{\pim}{\pi^{-}}
\newcommand{\piz}{\pi^{0}}
\newcommand{\az}{a_{0}(980)}
\newcommand{\azpi}{a^{\pm}_{0}(980)\pi^{\mp}}
\newcommand{\bone}{b_{1}(1235)}

\newcommand{\pipm}{\pi^{\pm}}

\newcommand{\jpsi}{J/\psi}
\newcommand{\ar}{\rightarrow}
\newcommand{\GeV}{GeV/$c^2$}
\newcommand{\MeV}{MeV/$c^2$}
\newcommand{\br}[1]{\mathcal{B}(#1)}

\newcommand{\ppbar}{p\bar{p}}

\begin{document}

\title{\Large \boldmath \bf Study of resonant structure around 1.8 \GeV~and $\eta(1405)$ in $\jpsi\ar\omega\eta\pip\pim$}

\author{
{\small M.~Ablikim$^{1}$, M.~N.~Achasov$^{5}$, D.~Alberto$^{41}$, Q.~An$^{39}$, Z.~H.~An$^{1}$, J.~Z.~Bai$^{1}$, R.~Baldini$^{19}$, Y.~Ban$^{25}$, J.~Becker$^{2}$, N.~Berger$^{1}$
% ...
\\
\vspace{0.2cm}
(BESIII Collaboration)\\
\vspace{0.2cm} {\it
$^{1}$ Institute of High Energy Physics, Beijing 100049, P. R. China\\
$^{2}$ Bochum Ruhr-University, 44780 Bochum, Germany\\
% ...
$^{5}$ G.I. Budker Institute of Nuclear Physics SB RAS (BINP), Novosibirsk 630090, Russia\\
% ...
$^{19}$ Laboratori Nazionali di Frascati - INFN, 00044 Frascati, Italy\\
% ...
$^{25}$ Peking University, Beijing 100871, P. R. China\\
% ...
$^{39}$ University of Science and Technology of China, Hefei 230026, P. R. China\\
% ...
$^{41}$ University of Turin and INFN, Turin, Italy\\
% ...
}}
\vspace{0.4cm} }


\collaboration{BES Collaboration}

\begin{abstract}
We present results of a study of the decay $\jpsi\ar\omega\eta\pip\pim$ using a sample of $(225.2\pm2.8)\times10^6$ $\jpsi$ events collected by the BESIII detector, and report the observation of a new process of $\jpsi\ar\omega X(1870)$ in which $X(1870)$ decays to $\azpi$. The statistical significance of this process is larger than $7.2\sigma$. Signals for $\jpsi\ar\omega f_1(1285)$ and $\jpsi\ar\omega\eta(1405)$ are also observed in $\eta\pip\pim$ spectrum, with statistical significances much larger than $10\sigma$.
\end{abstract}

\pacs{13.85.Hd, 25.75.Gz}

\maketitle

A new resonance, known as the $X(1835)$, was first observed in the $\etap\pip\pim$ mass spectrum of $\jpsi\ar\gamma\etap\pip\pim$ by BESII \cite{X1835_bes2} and subsequently confirmed with a much higher signal significance by BESIII \cite{X1835_bes3}. Several theoretical speculations have been proposed to interpret the nature of $X(1835)$, including the $\ppbar$ bound state \cite{X1835_th_ppbar1,X1835_th_ppbar2,X1835_th_ppbar3} that was first observed near the same mass in $\jpsi\ar\gamma\ppbar$ at BESII \cite{X1860_bes2} and confirmed by BESIII and CLEO \cite{X1860_bes3_CLEO}, a second radial excitation of the $\etap$ \cite{X1835_th_etap}, and a pseudo-scalar glueball \cite{X1835_th_glueball1,X1835_th_glueball2,X1835_th_glueball3}. In the lower mass region of the $\eta\pip\pim$ mass spectrum, around 1.4 \GeV, extensive studies \cite{eta1405_early1,eta1405_early2,eta1405_mark3} have established the existence of the $\eta(1405)$, which has also been suggested as a candidate for a pseudo-scalar glueball \cite{eta1405_L3}. Experimentally, the study of the production mechanism of the $X(1835)$ and $\eta(1405)$, {\it e.g.} searches for them in $\eta\pip\pim$ final states with other accompanying particles ($\omega$, $\phi$, etc.), are useful for clarifying their nature. In particular, the measurements of the production widths of these two states in hadronic decays of the $\jpsi$ and a comparison with corresponding measurements in $\jpsi$ radiative decay would provide important information about the glueball possibility \cite{jpsi_decay}.

In this letter, we present the results of a study of $\jpsi\ar\omega\eta\pip\pim$. A structure around 1.8$\sim$1.9 \GeV~in the $\eta\pip\pim$ mass spectrum is observed. This analysis is based on a sample of $(225.2\pm2.8)\times10^6$ $\jpsi$ events~\cite{jpsi_number} accumulated in the Beijing Spectrometer (BESIII)~\cite{bes3_design} operating at the Beijing Electron-Position Collider (BEPCII)~\cite{bepc2_design} at the Beijing Institute of High Energy Physics.


BEPCII is a double-ring $e^+e^-$ collider designed to provide $e^+e^-$ beams with a peak luminosity of $10^{33}~\rm{cm}^{-2}\rm{s}^{-1}$ at a beam current of 0.93 A. The cylindrical core of the BESIII detector consists of a helium-based main drift chamber (MDC), a plastic scintillator time-of-flight system (TOF), and a CsI(Tl) electromagnetic calorimeter (EMC) that are all enclosed in a superconducting solenoidal magnet providing a 1.0 T magnetic field. The solenoid is supported by an octagonal flux-return yoke with resistive plate counter muon identifier modules interleaved with steel. The acceptance of charged particles and photons is 93\% of 4$\pi$ steradians, and the charged-particle momentum and photon energy resolutions at 1~GeV are 0.5\% and 2.5\%, respectively. The BESIII detector is modeled with a Monte Carlo (MC) simulation based on \textsc{geant}{\footnotesize4} \cite{geant4_1,geant4_2}.


Charged tracks in the BESIII detector are reconstructed using track-induced signals in the MDC. To optimize the momentum measurement, we select tracks in the polar angle range $|\cos\theta|<0.93$ and require that they pass within $\pm20$ cm from the Interaction Point (IP) in the beam direction and within $\pm2$ cm of the beam line in the plane perpendicular to the beam. Four tracks with net charge zero are required, and all tracks are assumed to be pions.

Electromagnetic showers are reconstructed from clusters of energy deposits in the EMC. The energy deposited in nearby TOF counters is included to improve the reconstruction efficiency and energy
resolution. Showers identified as photon candidates must satisfy fiducial and shower-quality requirements, {\it i.e.} the showers in the barrel region $(|\cos \theta|<0.8)$ must have a minimum energy of 25 MeV, while those from the endcaps $(0.86<|\cos \theta|<0.92)$ must have at least 50 MeV. The showers in the angular range between the barrel and endcap are poorly reconstructed and excluded from the analysis. To suppress showers from charged particles, a photon must be separated by at least $10^\circ$ from the nearest charged track. The EMC cluster timing requirements are used to suppress electronic noise and energy deposits unrelated to the event.

In the reconstruction of $\jpsi\ar\omega\eta\pip\pim$, the $\omega$ is reconstructed via its $\pip\pim\piz$ mode, and the $\eta$ and $\piz$ are reconstructed from $\gamma\gamma$ pairs. The vertex of all final state particles must be consistent with the measured beam interaction point. The sum of the four-momenta of all particles is constrained to the known $\jpsi$ mass and the initial $e^+e^-$ three-momentum in the lab frame. The vertex and four-momentum kinematic fits are required to satisfy the quality requirements $\chi^2_{V}<100$ and $\chi^2_{4C}<50$, respectively. Further selections are based on the four-momenta from the kinematic fit. Photon pairs with an invariant mass satisfying $M_{\gamma\gamma}\in$ $(524,572)$ \MeV~or $(122,148)$ \MeV~are identified as $\eta$ or $\piz$ candidates. The $\eta\piz4\pi$ combination with minimum $\chi^2_{4C}$ is selected in the cases where more than one candidate satisfies the above requirements in an event. If there is more than one four-photon combination in the mass range of the $\eta$ and $\piz$, the assignment with the lowest value of $\chi_{\eta\piz}=\sqrt{P^2_{\eta}+P^2_{\piz}}$ is used, where
$P_{\eta/\piz}$ are the pulls defined as $P_{\eta/\piz}=\frac{M_{\gamma\gamma}-m_{\eta/\piz}}{\sigma_{\gamma\gamma}}$. Here $\sigma_{\gamma\gamma}$ is the $\eta/\piz$ mass resolution determined from data.

\begin{figure}[hbtp]
  \begin{center}
    \includegraphics[width=2.5in,height=2.5in]{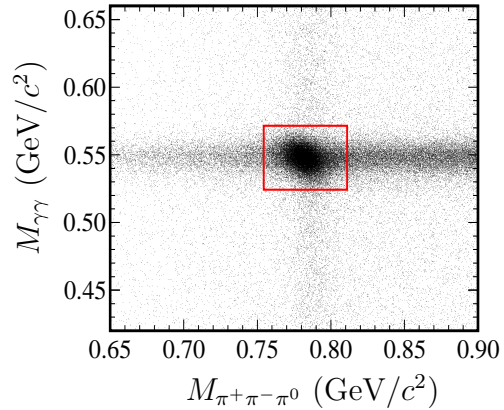}
    \put(-125,10){\large $M_{\pip\pim\piz}$ (\GeV)}
    \put(-190,65){\rotatebox{90}{\large $M_{\gamma\gamma}$ (\GeV)}}
    \caption{Scatter plot of $M_{\eta(\gamma\gamma)}$ versus
    $M_{\omega(\pip\pim\piz)}$. The rectangle in the middle shows
    the signal region defined as $|M_{\pip\pim\piz}-m_{\omega}|<$
    28 \MeV~and $|M_{\gamma\gamma}-m_{\eta}|<$ 24 \MeV.}
    \label{fig:mass_scatter}
  \end{center}
\end{figure}

After the application of the above requirements, the scatter plot of $M_{\eta(\gamma\gamma)}$ versus $M_{\omega(\pip\pim\piz)}$ (shown in Fig.~\ref{fig:mass_scatter}) shows a clear cluster in the $\jpsi\ar\omega\eta\pip\pim$ signal region denoted by the rectangle in the center of the plot.  To determine which $\pip\pim$ pair originates from the $\omega$, $|M_{\pip\pim\piz}-m_{\omega}|$ is minimized among the possible combinations of the selected charged pions, and required to
be less than 28 \MeV.

\begin{figure}[hbtp]
  \centering
    \includegraphics[width=1.6in,height=1.6in]{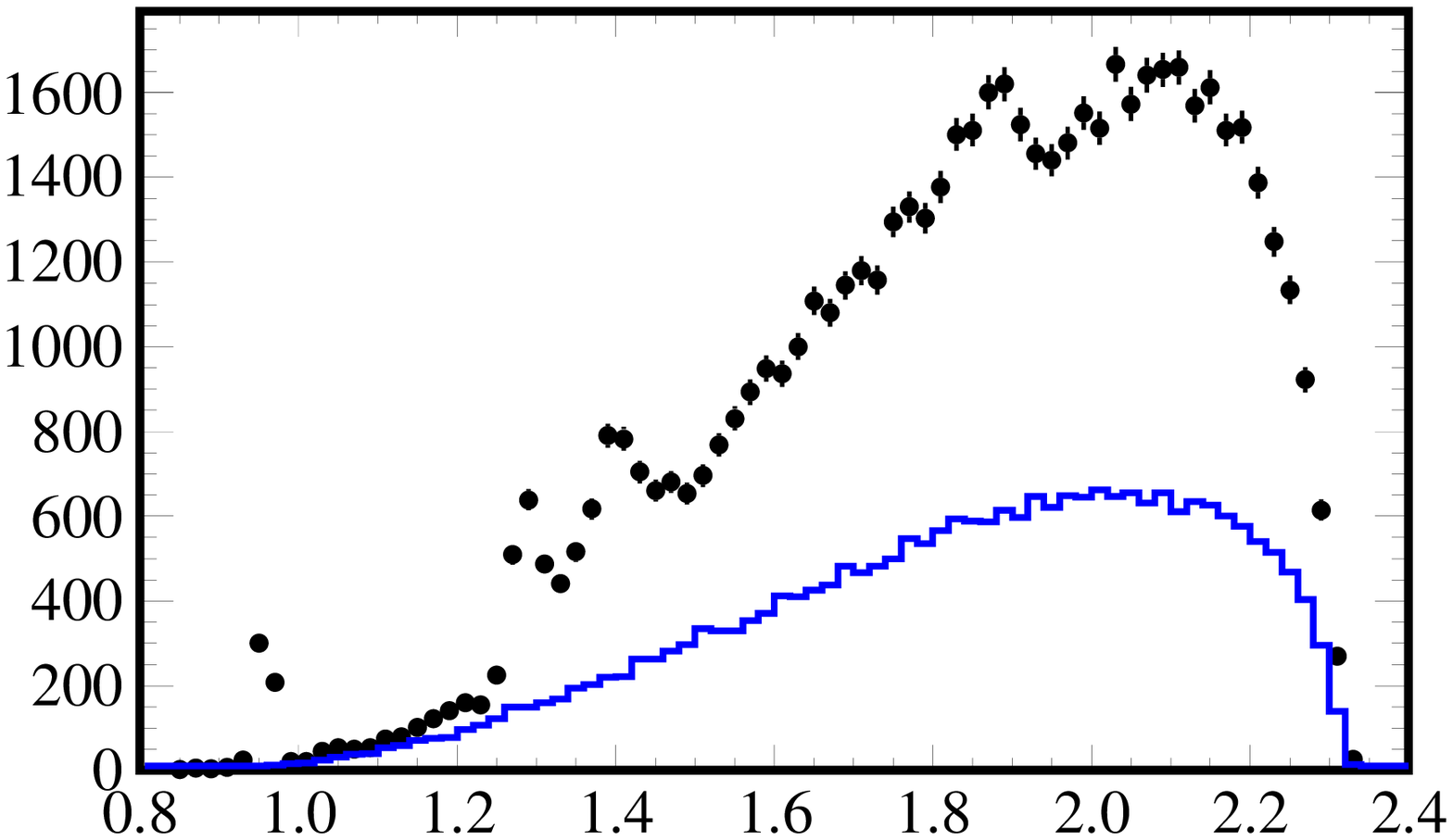}
    \includegraphics[width=1.6in,height=1.6in]{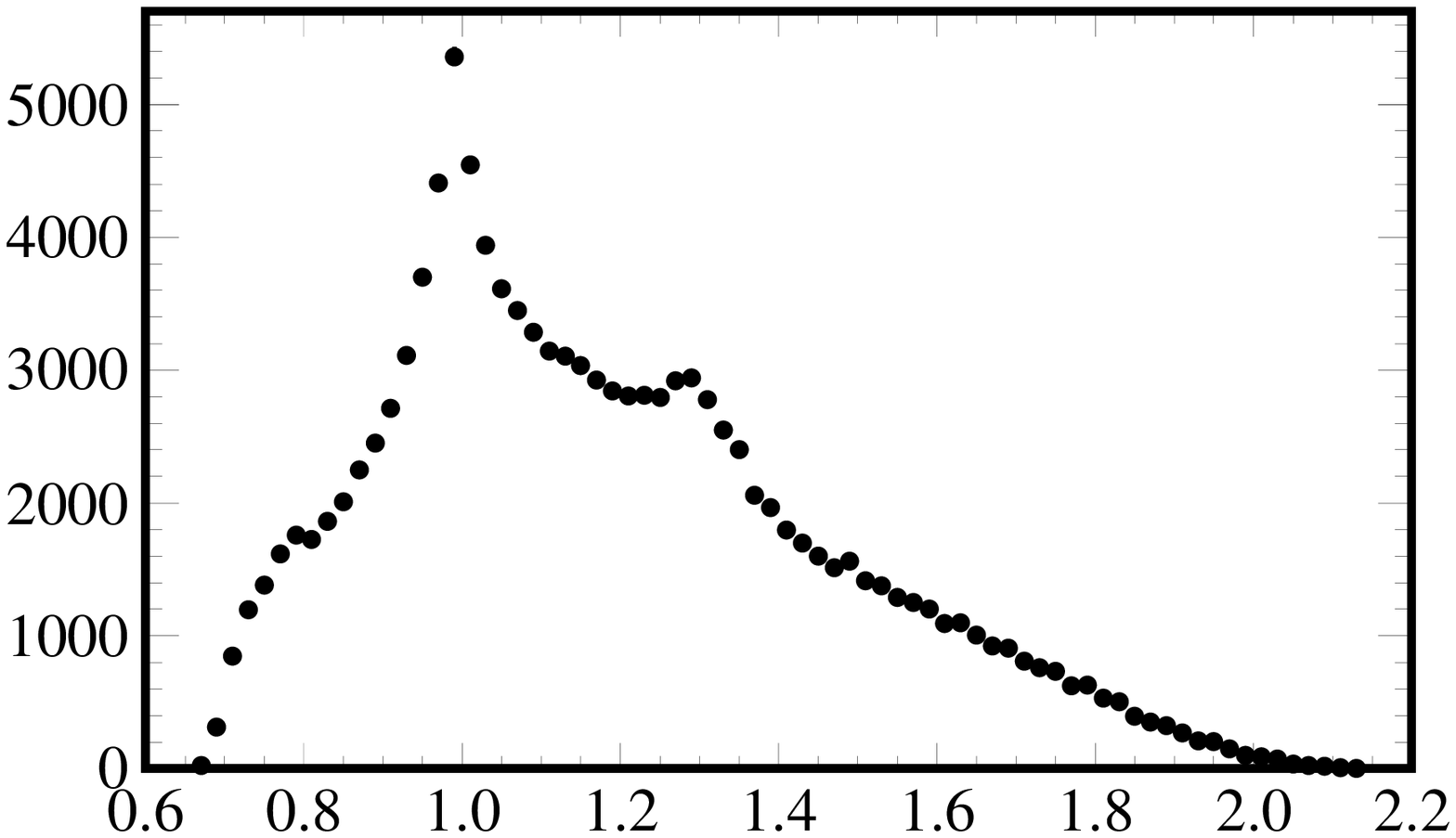}
    \put(-210,90) {(a)} \put(-92,90){(b)}
    \put(-210,5){$M_{\eta\pip\pim}$ (\GeV)}
    \put(-85,5){$M_{\eta\pipm}$ (\GeV)}
    \put(-240,25){\rotatebox{90}{Events / 20\MeV}}
    \newline
    \includegraphics[width=1.6in,height=1.6in]{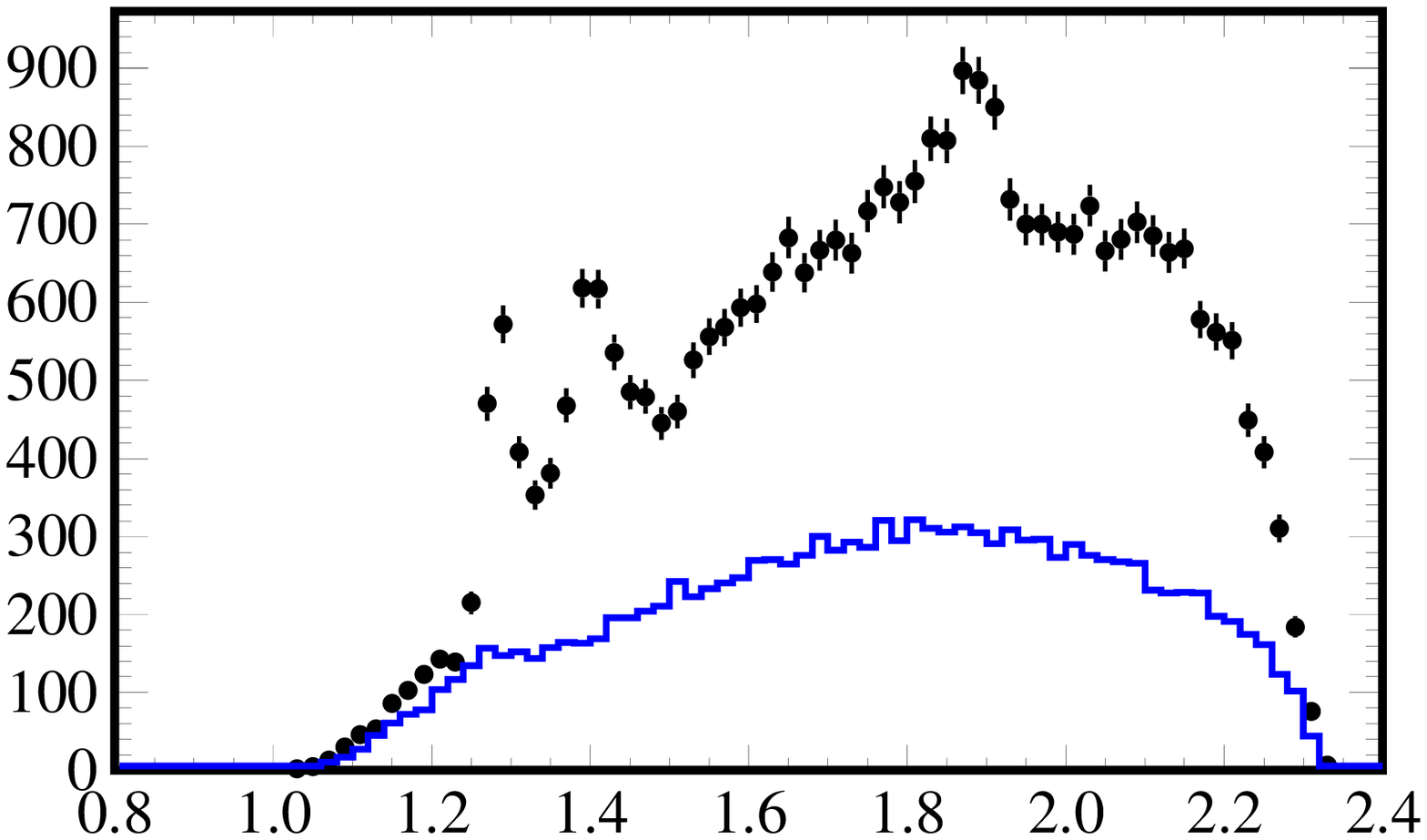}
    \includegraphics[width=1.6in,height=1.6in]{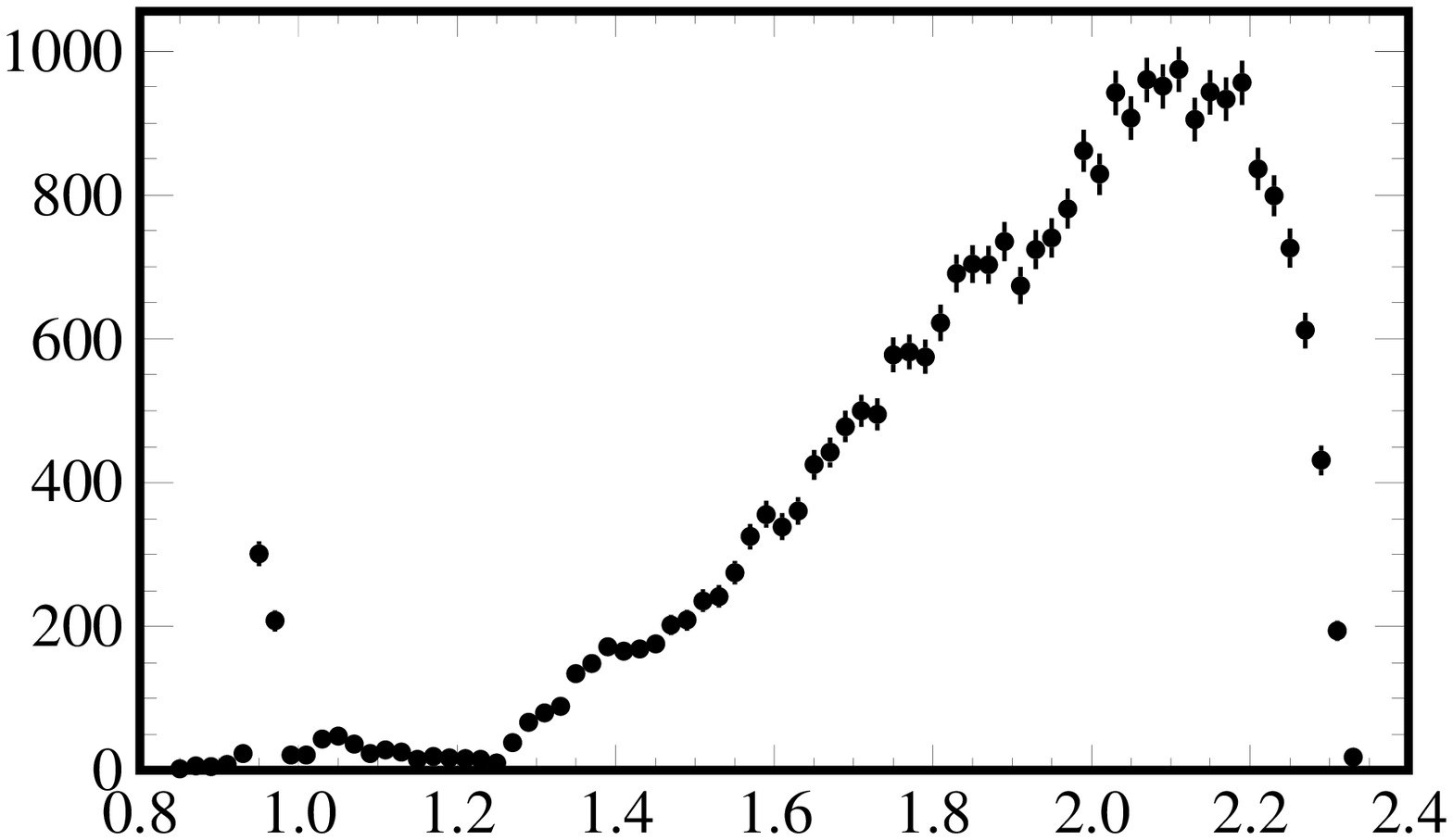}
    \put(-210,90) {(c)} \put(-92,90){(d)}
    \put(-210,5){$M_{\eta\pip\pim}$ (\GeV)}
    \put(-92,5){$M_{\eta\pip\pim}$ (\GeV)}
    \put(-240,25){\rotatebox{90}{Events / 20\MeV}}
  \caption{Invariant mass distributions for the selected candidate
  events: (a) and (b) are the invariant mass spectra of $\eta\pip\pim$
  and $\eta\pipm$ after the application of all the event selection
  criteria; (c) is the $\eta\pip\pim$ mass spectrum for events with an
  $\az$ in the $\eta\pipm$ final state; (d) is the $\eta\pip\pim$
  invariant mass distribution for events with no $\az$ in the
  $\eta\pipm$ system. The histograms in (a) and (c) are the
  phase-space MC events of $\jpsi\ar\omega\eta\pip\pim$ after the same
  event selection and with arbitrary normalization.}
  \label{fig:mass_specs}
\end{figure}

With all the selection criteria applied, the mass spectrum of $\eta\pip\pim$ is shown in Fig.~\ref{fig:mass_specs}(a). In the lower mass range, in addition to the well-known $\etap$ peak, two other structures are observed; these are inferred to be the $f_1(1285)$ and $\eta(1405)$ based on the fit results discussed below. There is an additional structure located around 1.87 \GeV~that we denote as $X(1870)$. The $\eta\pipm$ mass spectrum for these events shown in Fig.~\ref{fig:mass_specs}(b), reveals a strong $\az$ signal. The $\eta\pip\pim$ mass spectrum for events where either $M(\eta\pip)$ or $M(\eta\pim)$ is in a 100 \MeV~mass window centered on the $\az$ mass is shown in Fig.~\ref{fig:mass_specs}(c). The $\eta\pip\pim$ mass spectrum for events with both $M(\eta\pip)$ and $M(\eta\pim)$ outside the $a_0(980)$ signal region is shown in
Fig.~\ref{fig:mass_specs}(d). A comparison between of Fig.~\ref{fig:mass_specs}(c) and Fig.~\ref{fig:mass_specs}(d) indicates that the $f_1(1285)$, $\eta(1405)$ and $X(1870)$ all
primarily decay via the $a_0(980)\pi$ channel.

\begin{figure}[hbtp]
  \begin{center}
    \includegraphics[width=2.5in,height=2.5in]{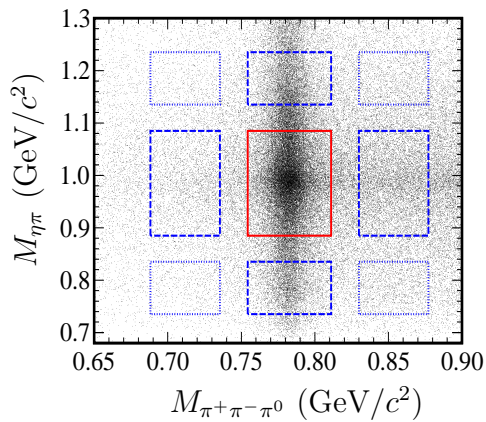}
    \put(-125,10){\large $M_{\pip\pim\piz}$ (\GeV)}
    \put(-185,65){\rotatebox{90}{\large $M_{\eta\pi}$ (\GeV)}}
    \caption{Definition of the signal and two-dimensional side bands.}
    \label{fig:mass_OmegaA0}
  \end{center}
\end{figure}

To ensure that the observed $f_1(1285)$, $\eta(1405)$ and the structure around 1.87 \GeV~originate from the process of $\jpsi\ar\omega\azpi$ rather than peaking backgrounds, potential background channels are studied using both data and MC samples. The non-$\omega$ and/or non-$\az$ processes are estimated by the events in the two-dimensional mass-sidebands of $\omega$ and $\az$ illustrated by the dashed and dotted boxes in Fig.~\ref{fig:mass_OmegaA0}, where the difference of the background shape in the signal region and side bands due to the varying phase space is taken into account by multiplying the $\eta\pip\pim$ mass distribution by a correction curve. In practice, this contribution is evaluated by the weighted-sums of horizontal and vertical side bands, with the entries in the diagonal side bands subtracted to compensate for the double counting of background components. The weighting factors for the events in the horizontal, vertical and the diagonal side bands are measured to be 0.48, 1.58 and 0.76, respectively, which are determined from the results of a two-dimensional fit to the mass spectrum of $M_{\omega}(\pip\pim\piz)$ versus $M_{\az}(\eta\pi)$. Here the two-dimensional Probability Density Functions (PDFs) for
$\jpsi\ar\omega\az\pi$, $\omega$ but non-$\az$, non-$\omega$ but $\az$, non-$\omega$ and non-$\az$ processes are constructed by the product of one-dimensional functions, where the resonant peaks are parametrized by Breit-Wigner functions and the non-resonant parts are described by floating polynomials. A MC sample of two million events of the phase space process of $\jpsi\ar\pip\pim\piz\eta\pip\pim$ is used to determine the correction curve to account for the varying phase space.

The background channel $\jpsi\ar\bone\az$, where the $\bone$ decays to $\omega\pi$ and $\az$ decays to $\eta\pi$, is studied by performing a two-dimensional fit to the $M(\omega\pi)$ versus $M(\eta\pi)$ mass distribution with two-dimensional PDFs defined in similar fashion. We also studied an inclusive MC sample of $2\times10^8$ $\jpsi$ decays generated according to the Particle Data Group (PDG) decay table and Lund-charm model \cite{lund}. The reliability of the background
estimation method described above is validated, and no background-induced peaks are observed around 1.87 \GeV.

\begin{figure}[hbtp]
  \begin{center}
    \includegraphics[width=2.5in,height=2.5in]{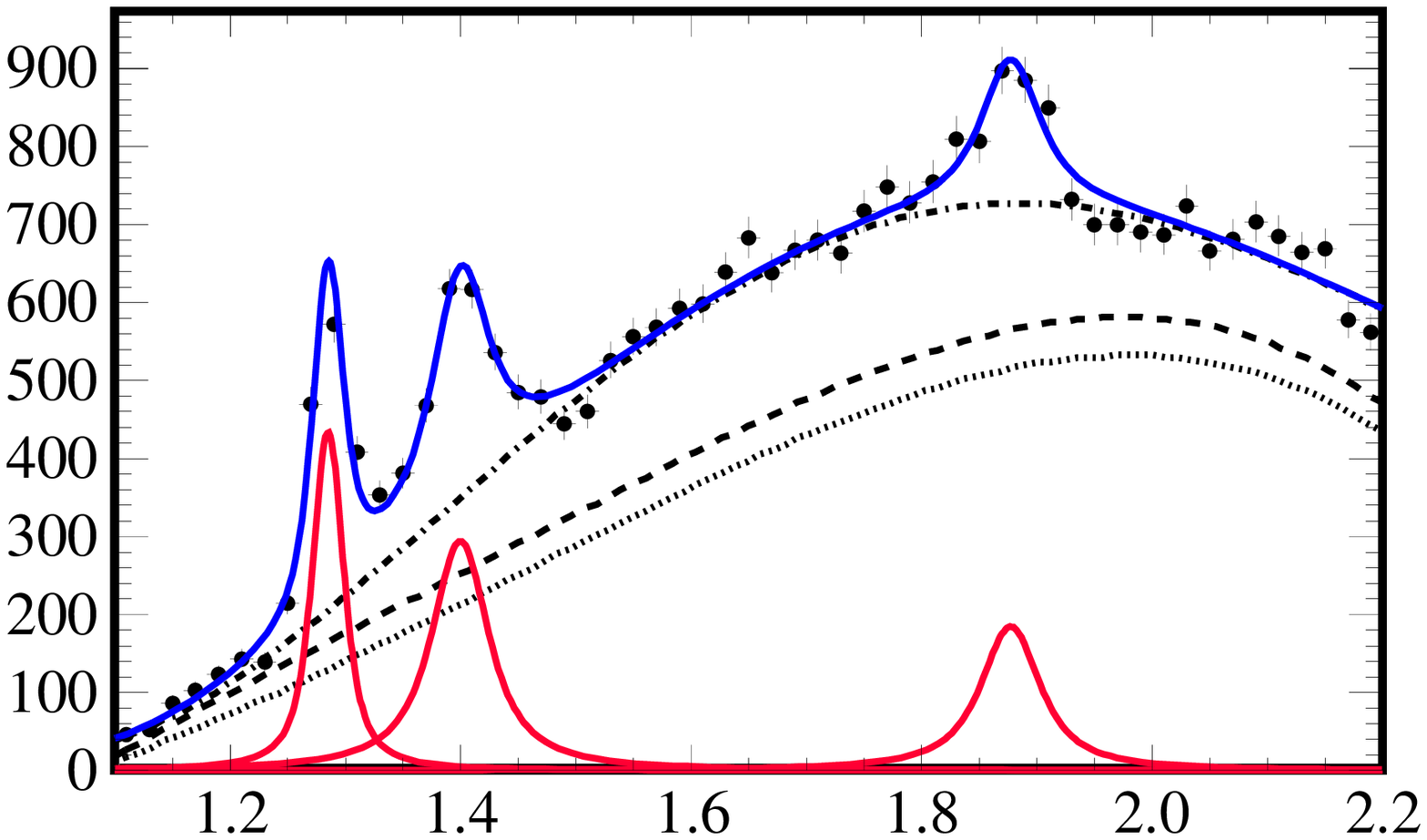}
    \put(-125,10){\large $M_{\eta\pip\pim}$ (\GeV)}
    \put(-185,55){\rotatebox{90}{\large Events / 20\MeV}}
    \caption{Results of fits to the $M(\eta\pip\pim)$ mass
     distribution for events with either the $\eta\pip$ or $\eta\pim$
     in the $\az$ mass window. The dotted curve shows the contribution
     of non-$\omega$ and/or non-$\az$ background, the dashed line also
     includes the contribution from $\jpsi\ar\bone\az$, and the
     dot-dashed curve indicates the total background with the
     non-resonant $\jpsi\ar\omega\azpi$ included. $\chi^2/d.o.f$ is
     1.27 for this fit.}
    \label{fig:mass_fit}
  \end{center}
\end{figure}

Figure \ref{fig:mass_fit} shows the results of a fit to the $\eta\pip\pim$ mass spectrum where either $\eta\pip$ or $\eta\pim$ are in the $\az$ mass window. Here the three signal peaks are
parametrized by efficiency-corrected Breit-Wigner functions convolved with a Gaussian resolution function, the width of which is determined from signal MC samples and fixed in the fit. The background consists of three components, namely contributions from non-$\omega$ and/or non-$\az$ processes, $\jpsi\ar\bone\az$ events, and non-resonant $\omega\az\pi$ processes. The background shapes and numbers of events for the non-$\omega$ and/or non-$\az$ processes are determined from
the events in the two-dimensional side bands as discussed above, and fixed in the fit. For the $\jpsi\ar\bone\az$ component of background, the background shape is fixed to that of the phase space MC samples whereas the number of events is extracted and fixed to the result of a two dimensional fit to the $\omega\pi$ versus $\eta\pi$ mass distributions. The contribution of the remaining non-resonant $\omega\az\pi$ process is described by a smooth floating polynomial function. The mass, width, and the product branching fractions obtained from the fit are summarized in Table \ref{tab:summary}. For the $f_1(1285)$ and $\eta(1405)$, the measured mass and width are in
agreement with PDG values \cite{PDG}. A conservative estimate of the statistical significance of the $X(1870)$ signal is determined by the lower limit in the change of $-2\ln\mathcal{L}$ obtained from the fits with and without the assumption of a $X(1870)$. With all the factors in the fit varied, the smallest change in the $-2\ln\mathcal{L}$ is 60.1, corresponding to a significance of $7.2\sigma$. The same procedure is applied to the $f_1(1285)$ and $\eta(1405)$ signals, and the significances are determined to be much higher than 10 $\sigma$.

%
%
%
%

\begin{table}[hbtp]
  \caption{Summary of measurements of the mass, width and the product
  branching fraction of $\br{\jpsi\ar\omega X}
  \times\br{X\ar\azpi}\times\br{a_{0}^{\pm}(980)\ar\eta\pipm}$ where
  $X$ represents $f_1(1285)$, $\eta(1405)$ and $X(1870)$. Here the
  first errors are statistical and the second ones are systematic.}
  \label{tab:summary}\scriptsize
  \begin{center}
     \renewcommand{\arraystretch}{1.5}
     \begin{tabular}{c|ccc}
        \hline\hline
        Resonance     & Mass (\MeV)          & Width (\MeV)       & $\mathcal{B}$ ($10^{-4}$) \\ \hline

        $f_{1}(1285)$ & $1285.1\pm1.0^{+1.6}_{-0.3}$ & $22.0\pm3.1^{+2.0}_{-1.5}$ & $1.25\pm0.10^{+0.19}_{-0.20}$ \\

        $\eta(1405)$  & $1399.8\pm2.2^{+2.8}_{-0.1}$ & $52.8\pm7.6^{+0.1}_{-7.6}$ & $1.89\pm0.21^{+0.21}_{-0.23}$ \\

        $X(1870)$     & $1877.3\pm6.3^{+3.4}_{-7.4}$ & $57\pm12^{+19}_{-4}$       & $1.50\pm0.26^{+0.72}_{-0.36}$ \\

        \hline\hline
      \end{tabular}
  \end{center}
\end{table}


The systematic errors on the measurement of the mass and width parameters are primarily due to the uncertainty in the mass spectrum fitting. In detail, the fit range, background estimation method, number of background events, and the background parametrization are varied to decide the uncertainty from the background estimation and fitting as a whole. We also include the systematic errors determined from the input/output checks based on the analysis of full-reconstructed MC samples, in which the input parameters are set according to the final results and the background is represented by the background channels seen in the inclusive MC sample. For systematic errors originating from the potential structure around 2.2~\GeV~and the multiple-event candidate selection, we re-fit the mass spectrum of $\eta\pip\pim$ with the inclusion of a $X(2120)$ resonance as recently reported by BESIII \cite{X1835_bes3} in the decay channel of $\jpsi\ar\gamma\etap\pip\pim$, and all the valid multiple-entry candidates kept in order to estimate the uncertainty due to these two sources, respectively. With the numbers from all the sources above combined quadratically, the systematic errors on the mass and width parameters are determined as shown in Table~\ref{tab:summary}.

The systematic errors on the branching fraction measurements are also subject to errors of the number of $\jpsi$ events \cite{jpsi_number}, the intermediate branching fractions \cite{PDG}, the data-MC difference in the $\pi$ tracking efficiency, the photon detection efficiency, the kinematic fit, the signal selection efficiency of $\eta/\piz$, the simulation of the line shape of $\az$ \cite{a0_flatte}, and the angular distributions due to different possible spin-parity hypotheses. Combined in quadrature with the influence from the mass spectrum fitting, the systematic errors on the product branching fraction for the $f_1(1285)$, $\eta(1405)$ and $X(1870)$ are summarized in Table~\ref{tab:summary}.


In summary, we present a study of the $\jpsi\ar\omega\eta\pip\pim$ decay channel and report the first observation of a new process of $\jpsi\ar\omega X(1870)$ in which $X(1870)$ decays to $\azpi$, with the signal significance estimated to be $7.2\sigma$. In the lower mass region of $\eta\pip\pim$ mass spectrum, the $f_1(1285)$ and $\eta(1405)$ are also clearly observed with statistical significances much larger than $10\sigma$. The measurements of the mass, width, and
product branching fraction of $\br{\jpsi\ar\omega X}\times\br{X\ar\azpi}\times\br{a_{0}^{\pm}(980)\ar\eta\pipm}$ for the three resonant structures are summarized in Table \ref{tab:summary}, wherein the branching fractions for the $f_1(1285)$ and $\eta(1405)$ are measured for the first time. Whether the resonant structure of $X(1870)$ is due to the $X(1835)$, the $\eta_2(1870)$, an interference of both, or a new resonance still needs further study such as a partial wave analysis that will be possible with the larger $\jpsi$ data sample that is anticipated in future runs of the BESIII experiment. For the $\eta(1405)$, the product branching fraction of its hadronic production is measured to be smaller than that for its production in the radiative $\jpsi$ decays \cite{PDG}.


We thank the accelerator group and computer staff of IHEP for their effort in producing beams and processing data. We are grateful for support from our institutes and universities and from these agencies: Ministry of Science and Technology of China, National Natural Science Foundation of China, Chinese Academy of Sciences, Istituto Nazionale di Fisica Nucleare, Russian Foundation for Basic Research, Russian Academy of Science (Siberian branch), U.S. Department of Energy, and National Research Foundation of Korea.



\begin{thebibliography}{99}

\bibitem{X1835_bes2} M. Ablikim {\em et al.} (BES Collaboration), Phys. Rev. Lett. {\bf95}, 262001 (2005).

\bibitem{X1835_bes3} M. Ablikim {\em et al.} (BESIII Collaboration), Phys. Rev. Lett. {\bf106}, 072002 (2011).

\bibitem{X1835_th_ppbar1} S. L. Zhu and C. S. Gao, Commun. Theor. Phys. {\bf46}, 291 (2006).

\bibitem{X1835_th_ppbar2} G. J. Ding and M. L. Yan, Phys. Rev. C {\bf72}, 015208 (2005).

\bibitem{X1835_th_ppbar3} J. P. Dedonder {\em et al.}, Phys. Rev. C {\bf80}, 045207 (2009).

\bibitem{X1860_bes2} J. Z. Bai {\em et al.} (BES Collaboration), Phys. Rev. Lett. {\bf91}, 022001 (2003).

\bibitem{X1860_bes3_CLEO} M. Ablikim {\em et al.} (BESIII Collaboration), Chinese Physics C {\bf34}, 421-426 (2010); J. P. Alexander {\em et al.} (CLEO Collaboration), Phys. Rev. D {\bf82}, 092002 (2010).

\bibitem{X1835_th_etap} T. Huang and S. L. Zhu, Phys. Rev. D {\bf73}, 014023 (2006).

\bibitem{X1835_th_glueball1} G. Hao, C. F. Qiao and A. L. Zhang, Phys. Lett. B {\bf642}, 53 (2006).

\bibitem{X1835_th_glueball2} B. A. Li, Phys. Rev. D {\bf74}, 034019 (2006).

\bibitem{X1835_th_glueball3} N. Kochelev and D. P. Min, Phys. Lett. B {\bf633}, 283 (2006).

\bibitem{eta1405_early1} D. L. Scharre {\em et al.}, Phys. Lett. {\bf97B}, 329 (1980)

\bibitem{eta1405_early2} C. Edwards {\em et al.}, Phys. Rev. Lett. {\bf49}, 259 (1982)

\bibitem{eta1405_mark3} T. Bolton {\em et al.} (Mark III Collaboration), Phys. Rev. Lett. {\bf69}, 1328 (1992).

\bibitem{eta1405_L3} M. Acciarri {\em et al.} (L3 Collaboration), Phys. Lett. B {\bf501}, 1 (2001).



\bibitem{jpsi_decay} L. K\"{o}pke and N. Wermes, Physics Reports {\bf 174}, Nos. 2 \& 3, 67-227 (1989).

\bibitem{jpsi_number} M. Ablikim {\em et al.} (BESIII Collaboration), Phys. Rev. D {\bf83}, 012003 (2011).


\bibitem{bes3_design} M. Ablikim {\it et al.} (BESIII Collaboration), Nucl. Instrum. Methods Phys. Res. A {\bf 614}, 345 (2010).

\bibitem{bepc2_design} J. Z. Bai {\it et al.} (BES Collaboration), Nucl. Instrum. Methods Phys. Res. A {\bf 344}, 319 (2001).

\bibitem{geant4_1} S. Agostinelli {\it et al.} (\textsc{geant}{\footnotesize4} Collaboration), Nucl. Instrum. Methods Phys. Res. A {\bf 506}, 250 (2003).

\bibitem{geant4_2} J. Allison {\it et al.}, IEEE Trans. Nucl. Sci. {\bf 53}, 270 (2006).

\bibitem{lund} R. G. Ping, Chinese Phys. C {\bf 32}, 599 (2008).

\bibitem{PDG} K. Nakamura {\it et al.}  (Particle Data Group), J. Phys. G {\bf 37}, 075021 (2010).

\bibitem{a0_flatte} Jia-Jun Wu and B. S. Zou, Phys. Rev. D {\bf 78}, 074017 (2008)

\end{thebibliography}
\end{document}